\documentstyle[preprint,prb,aps]{revtex}

\begin{document}
\draft
\title{The role of the buildup oscillations on the speed of resonant tunneling
diodes}
\author{Roberto Romo \cite{byline} and Jorge Villavicencio}
\address{Facultad de Ciencias,Universidad Aut\'onoma de Baja California\\
Apartado Postal 1880, 22800 Ensenada, Baja California, M\'exico}
\date{\today}
\maketitle

\begin{abstract}
The fastest tunneling response in double barrier resonant structures is
investigated by considering explicit analytic solutions of the time
dependent Schr\"{o}dinger equation. For cutoff initial plane waves, we find that the earliest tunneling
events consist on the emission of a series of propagating pulses of the
probability density governed by the buildup oscillations in the quantum
well. We show that the fastest tunneling response comes from the contribution of 
incident carriers at energies different from resonance, and that its relevant time scale is given by $\tau _r=\pi \hbar /\left| E-\varepsilon \right| $, where $\varepsilon $ is the resonance energy and $E$ is the incidence energy.
\end{abstract}


\widetext
\mediumtext
\narrowtext

\narrowtext
The transport mechanism in resonant tunneling diodes (RTD) and the relevant
time scale for its fastest tunneling response have been the subject of
intense investigation. \cite{intinv,soll83,RAzbel,luryi} In particular, the
results of the high-frequency tunneling experiments \cite{soll83}
demonstrated the possibility that the charge transport may occur at time
scales even shorter than the lifetime of the quasibound state of the system.
Most of the theoretical efforts to estimate the relevant time scales for the
tunneling process have been based on stationary approaches;\cite
{RAzbel,luryi} however, it has been widely recognized \cite
{soll83,RAzbel,Sakakis} that the analytic solution of the time-dependent
Schr\"{o}dinger equation (TDSE) provides the most reliable way to tackle
this fundamental issue. Although there is in the literature a consensus that
the buildup dynamics determines the ultimate speed of RTD's, the way in
which the buildup governs the emission of transmitted particles still needs
elucidation.

In this letter we investigate the earliest tunneling events in double
barrier resonant structures and its relation to the dynamics of the buildup
process in the quantum well, based on the analytic solution of the TDSE.
Starting from the formal solution, we show that the earliest evidence of
transmitted carriers to the right edge of the structure ($x\geq L$) consists
on the emission of a series of propagating pulses, which are governed by the
buildup oscillations\cite{apl} in the quantum well.

The initial condition used here is inspired on experimental situations in
which the tunneling process begins almost abruptly. It is represented by a
cutoff planewave $\Psi \left( x,k;t=0\right) =\Theta (-x)(e^{ikx}-e^{-ikx})$%
\ impinging on a shutter\cite{ci} placed at $x=0$, just at the left edge of
a DB structure. The tunneling process begins with the instantaneous opening
of the shutter at\ $t=0$, enabling the incoming wave to interact with the
potential, giving rise to both the buildup in the quantum well and the
electronic transport to the right of the DB.{\it \ }For the above initial
condition, the solutions $\Psi ^{i}(x,k;t)$ for the internal region ($0\leq
x\leq L$), and $\Psi ^{e}(x,k;t)$ for the external region ($x>L$), are
respectively: \cite{formalismo} 
\begin{eqnarray}
\Psi ^{i} &=&\phi _{k}M(y_{k}^{i})-\phi
_{-k}M(y_{-k}^{i})-i\sum\limits_{n=-\infty }^{\infty }\phi
_{n}M(y_{k_{n}}^{i}),  \label{Psiint} \\
\Psi ^{e} &=&T_{k}M(y_{k}^{e})-T_{-k}M(y_{-k}^{e})-i\sum\limits_{n=-\infty
}^{\infty }T_{n}M(y_{k_{n}}^{e}).  \label{Psiext}
\end{eqnarray}
Here $\phi _{k}\equiv \phi (x,k)$ is the stationary wave function, $%
T_{k}\equiv t(k)$ is the transmission amplitude, and the factors $\phi
_{n}(x,k)\equiv 2ku_{n}(0)u_{n}(x)/(k^{2}-k_{n}^{2})$ and $T_{n}=\phi
_{n}(L,k)\exp (-ik_{n}L)$ are given in terms of the resonant eigenfunctions $%
u_{n}(x)$ with complex eigenvalues $k_{n}=a_{n}-ib_{n}$ ($a_{n},b_{n}>0$).
The index $n$ runs over the complex poles $k_{n}$ distributed in the third
and fourth quadrants in the complex $k$-plane. The $M^{\prime }s$ are the
Moshinsky functions,\cite{PRA97} with arguments $y_{q}^{e}(x,t)=e^{-i\pi
/4}(m/2\hbar t)^{1/2}[x-\hbar qt/m]$, and $y_{q}^{i}=y_{q}^{e}(0,t)$, where $%
q$ stands either for $\pm k$ or $k_{\pm n}$.{\small \ }For the particular
case of DB systems with isolated resonances i.e. $|\varepsilon _{n\pm
1}-\varepsilon _{n}|\ll \Gamma _{n}$, one term is sufficient;\cite{apl,prbrc}
most of the DB diodes with typical potential parameters fall into this
category.

The emission of the earliest transmitted particles is described by the
external probability density; this is illustrated in Fig. \ref{fig1} in
which we plot the normalized probability density $\left| \Psi
^{e}(x,k;t)/T_{k}\right| ^{2}$\ as a function of $x$\ for fixed values of
time chosen in this example\cite{parameters} as $t=2$ and $10$ ps. For
off-resonance incidence (solid lines), the propagation of an oscillatory
structure with a sharp defined wavefront is clearly appreciated; to
illustrate that this wavefront travels with approximately the classical
speed $v=(2E/m)^{1/2}$, the arrow indicates the position $x=vt$, for $t=10$
ps. Note that $\left| \Psi ^{e}\right| ^{2}$ oscillates around the
transmission coefficient $\left| T_{k}\right| ^{2}$, which is the expected
asymptotic value as $t\rightarrow \infty $ {\em at }$x=L${\em .}

In order to analyze the tunneling mechanism at its earliest stages, we
present in Fig. \ref{fig2} a series of snapshots of this process at early
times and their correlation with the periodic buildup oscillations observed
in the quantum well. In part (a) we illustrate the buildup during this
transient at the particular times in which the buildup in the well reaches
its maxima and minima, as it oscillates around the stationary value (dotted
line). Panel (b) exhibits the birth of the first pulses for the same
sequence of times chosen in (a). The correspondence between the buildup
oscillations and the successive emission of pulses is evident from the
figure. We see that upon completion of one buildup cycle, a pulse is fired;
in this way for example, at $2.4$ ps, three cycles have been completed and
up to this stage three pulses have already been emitted.

In order to find analytically how the buildup governs the fastest tunneling
response and its relevant time scales, we shall exploit the analytical
properties of the solutions. As shown in a recent paper,\cite{apl} the
probability density at any position of the internal region, may be described
by the simple formula

\begin{equation}
\left| \Psi ^i/\phi \right| ^2=1+e^{-\Gamma _nt/\hbar }-2e^{-\Gamma
_nt/2\hbar }\cos \left[ \omega _nt\right] ,  \label{Psicos}
\end{equation}
where $\omega _n=\left| E-\varepsilon _n\right| /\hbar $. The oscillatory
function of the above formula gives the periodicity of the buildup
oscillations depicted in Fig. \ref{fig2} (a). As we shall show below this
dynamical behavior also manifests itself outside the structure, and is the
key mechanism behind the observed synchronization between the buildup
oscillations and the pulse emission. Although the external solution, Eq. (%
\ref{Psiext}), is more complex than $\Psi ^i$ since the $M$ functions
involved in $\Psi ^e$ depend on both position and time, a simple expression
for $\left| \Psi ^e/T_k\right| ^2$ can be obtained. We start by considering
the one-level formula for the external solution, 
\begin{eqnarray}
\Psi ^e &=&T_kM(y_k^e)-T_{-k}M(y_{-k}^e)-iT_nM(y_{k_n}^e)-  \nonumber \\
&&iT_{-n}M(y_{k_{-n}}^e).  \label{one-res}
\end{eqnarray}

From the one level expression for $\phi (L,k)$,\cite{prbrc} the transmission
amplitude $T_k$ can be written as $T_k=2ik\exp
(-ikL)u_n(0)u_n(L)/(k^2-k_n^2) $; thus the factors $iT_n$ and $iT_{-n}$ in
the above expression can be readily identified as $iT_n=T_k\exp [i(k-k_n)L]$
and $iT_{-n}=-T_k^{*}\exp [-i(k-k_n^{*})L]$. Using the symmetry relation\cite
{PRA97} for the $M$ functions, $M(y_q^e)=\exp (y_q^{e2})-M(-y_q^e)$, in $%
M(y_k^e)$ and $M(y_{k_n}^e)$, we obtain a suitable representation for $\Psi
^e$ consisting on exponential terms and $M$ functions with arguments of the
type $y_q^e$ with $q=-k,$ $-k_n,$ $-k_n^{*}$, which have vanishingly small
contributions to the solution. In fact, for a fixed value of the position $%
x_f$, there exists a time interval starting from $t\gtrsim (mx_f/\hbar k)$,
governed exclusively by the exponential terms mentioned above. From these
considerations, a simple formula for $\left| \Psi ^e/T_k\right| ^2$ can be
obtained, namely,

\begin{eqnarray}
\left| \Psi ^e/T_k\right| ^2 &=&1+e^{-\Gamma _nt/\hbar
}e^{2b_n(x_f-L)}-2e^{b_n(x_f-L)}e^{-\Gamma _nt/2\hbar }  \nonumber \\
&&\times \cos \left[ (a_n-k)(x_f-L)+\omega _nt\right] ,  \label{lamdae}
\end{eqnarray}
valid for $t\gtrsim (mx_f/\hbar k)$. By comparison with Eq. (\ref{Psicos}),
we note that $\left| \Psi ^e/T_k\right| ^2$ and $\left| \Psi ^i/\phi \right|
^2$ have similar time dependence. In fact, at $x_f=L$, (\ref{lamdae})
reduces to

\begin{equation}
\left| \Psi ^{e}/T_{k}\right| ^{2}=1+e^{-\Gamma _{n}t/\hbar }-2e^{-\Gamma
_{n}t/2\hbar }\cos \left[ \omega _{n}t\right] ,  \label{ext-cos}
\end{equation}
which coincides exactly with Eq. (\ref{Psicos}). This analytical result
confirms the qualitative discussion of Fig. \ref{fig2}, that the buildup
oscillations in the quantum well govern the emission of the propagating
pulses at $x=L$. Therefore, both the buildup oscillations and the emitted
pulses are characterized by the same time scales. It\ is straightforward to
see from Eq. (\ref{Psicos}) that the maxima of the buildup oscillations
occur approximately at the time scales $\tau _{m}$\ given by 
\begin{equation}
\tau _{m}=\frac{(2m-1)\pi \hbar }{\Delta E_{n}},\qquad m=1,2,3,...
\label{scales}
\end{equation}
where $\Delta E_{n}\equiv \left| E-\varepsilon _{n}\right| $\ measures the
deviation of the incidence energy $E$\ from resonance $\varepsilon _{n}$.
Note that the same result is obtained for incidence above and below
resonance, $E=\varepsilon _{n}\pm \Delta E_{n}$, in view that Eq. (\ref
{ext-cos}) is irrespective to this choice. According to Eq. (\ref{ext-cos}),
the intensity of the pulses reaches maximum values at $x=L$ also at these
time scales. The formation and emission of the first of such pulses
constitutes the earliest evidence of carrier presence at the transmitted
region; thus, the relevant time scale for the fastest tunneling response
corresponds to $m=1$, and is simply given by 
\begin{equation}
\tau _{r}=\frac{\hbar \pi }{\Delta E}.  \label{tiempo}
\end{equation}

With regard to the times scales involved in the tunneling process, it is
important to emphasize the differences between the response time $\tau _r$
and other relevant time scales, such as the lifetime, $\tau _l=\hbar /\Gamma 
$, and the {\it buildup time,} $\tau _b$. In what follows we shall discuss
these differences and their implications.

The expressions for $\tau _r$ and $\tau _l$ are similar since they are both
inversely proportional to a certain energy width. However despite of this
resemblance, there is a fundamental difference: on the one hand the lifetime
is an {\it intrinsic} property of the resonant structure since it depends
exclusively on the system parameters through the resonance width $\Gamma $;
on the other hand, the time scale $\tau _r$ is not an intrinsic property of
the system since it takes into account external information, namely, the
energy of the incident carriers through the ``off-resonance width'', $\Delta
E$. This difference is crucial for understanding the ultrafast response in
DB diodes, and an important consequence is that depending on $\Delta E$, the
response time $\tau _r$ may be greater or even shorter than $\tau _l$; in
fact, it is easy to see from Eq. (\ref{tiempo}) that the condition for $\tau
_r<\tau _l$\ is simply $\Delta E>\pi \Gamma $. This is illustrated in Fig. 
\ref{fig3}, in which we compare the positions of the main peak of the plots
of $\left| \Psi ^e/T_k\right| ^2$ as a function of time and fixed position
at $x=L$, using two different values of $\Delta E$.

The buildup time $\tau _{b}$ is the duration of the transient regime in
which the time-dependent probability density $\left| \Psi ^{i}\right| ^{2}$
reaches its level-off given by the stationary probability density, $\left|
\phi \right| ^{2}$. As shown in a recent work,\cite{apl} such a transient is
of approximately ten lifetimes, for any DB structure with isolated
resonances; see Fig. 2 (b) of Ref. 6. As shown in the previous paragraph, $%
\tau _{r}$ can be shorter than $\tau _{l}$ and consequently than $\tau _{b},$
which means that several buildup oscillations (and emitted pulses) may occur
before $\tau _{b}$. This is clearly illustrated in Fig. \ref{fig2} where for
example at $t=2.4$\ ps{\em \ }($\approx 0.38\tau _{b}$) significant evidence
of transmitted carriers at the right of the structure is appreciated; in
fact, up to this time the head of the transmitted wave has already traveled
more than $10^{3}$\ nm, despite the fact that the buildup in the quantum
well has not been fully established.

The above results emphasize on the importance of the off-resonant carriers
in ultrafast tunneling, which play an important role in typical tunneling
experiment with RTD's,\cite{soll83} where the energies of the available
carriers are distributed within a finite interval $0<E<E_{F}$, where $E_{F}$
is the Fermi energy of the emitter. In fact, although off-resonance carriers
have a smaller transmission coefficient, the main contribution to the
tunneling current comes from energies near resonance rather than from the
resonance alone (i.e. the Tsu-Esaki formula involves an integration over the
whole Fermi interval, $0<E<E_{F}$). As an implication of our analysis, when
the resonance $\varepsilon $ is immersed into the Fermi sea and the
resonance width $\Gamma $ is such that $E_{F}>\pi \Gamma $, there will be
carriers of the Fermi interval fulfilling $\Delta E>\pi \Gamma $ and hence
contributing with tunneling responses faster than the lifetime. The above
situation should be present in tunneling experiments on DB systems in which $%
E_{F}\gg \pi \Gamma $.

We conclude our discussion with the following remarks: (i) Since the
tunneling dynamics critically depends on the shape of the incoming wave,
whenever one deals with the problem of the response time, one must clearly
specify to what initial condition it corresponds. In this respect, we stress
that the response time derived here corresponds to cutoff initial plane
waves. (ii) In view that plane waves are the ``building blocks'' of
wavefunctions with more general shapes, the results obtained here may also
give insight into the more intricate behavior expected for the case of
incoming wavepackets.\cite{hauge} (iii) The emphasis of the present study is
the exploration of genuine quantum dynamical effects occurring in the
transient regime; in particular, we have shown that the earliest tunneling
events in a DB structure consist on the firing of a series of propagating
pulses whose periodic emission at the right edge of the structure is
governed by the buildup oscillations in the quantum well. (iv) The
importance of the off-resonance incident carriers on the response time is
demonstrated within a purely coherent picture, and a closed formula (valid
near resonance) for the corresponding time scale has been derived from the
analytic solution.

The authors acknowledge financial support from Conacyt, M\'{e}xico, through
Contract No. 431100-5-32082E. The authors are also grateful with G. Garc\'{i}%
a-Calder\'{o}n for useful discussions.

\begin{figure}[tbp]
\caption{ Plot of $\left| \Psi ^e\left( x,k;t\right) /t(k)\right| ^2$ as a
function of the distance $x$ for two fixed values of time $t=2$, $10$ ps
(solid lines). The incidence energy is below resonance $E=\protect\varepsilon
-\Delta E=74.97$ $meV$ ($\Delta E=5\Gamma $); for comparison to the
classical propagation, the arrow indicates the position $x=vt$ for $t=10$
ps. The special case of incidence at resonance $E=80.11$ $meV$ is also
included for comparison (dashed lines).}
\label{fig1}
\end{figure}

\begin{figure}[tbp]
\caption{ Early times of the tunneling process in a DB structure. Part (a)
illustrates the buildup oscillations inside the structure (solid line)
around the stationary probability density $|\protect\phi |^2$ (dotted line);
the potential barriers are schematically represented by the dashed-dotted
lines. For the same sequence of times, part (b) shows the birth and emission
of the first pulses which are the earliest tunneling events occurring in the
structure. Note the synchronization with the buildup dynamics: for each
buildup oscillation, a pulse is emitted.}
\label{fig2}
\end{figure}

\begin{figure}[tbp]
\caption{ Plots of $\left| \Psi^e \left(x,k;t\right) /t(k)\right| ^2$ as a
function of time at the fixed position $x=L=15 $ nm, for two values of $%
\Delta E$. We see that depending on the value of $\Delta E$, the response
time $\protect\tau_r $ may be greater or shorter than the lifetime of the
quasibound state.}
\label{fig3}
\end{figure}


\begin{references}
\bibitem{parameters}  The parameters of the symmetrical DB structure are:
barrier heights $V_{0}=0.23$ eV, barrier widths $b_{0}=5$ nm, well width $%
\omega _{0}=5$ nm, and effective mass of the electron $m=0.067m_{e}$; the
position and width of the first resonance are respectively, $\varepsilon
=80.11$ meV, and $\Gamma =1.033$ meV. We have dropped here the subindex $n$
since we are dealing with the first resonance.
\end{references}
\end{document}